# Artificial Intelligence for Prediction of Climate Extremes: State of the art, challenges and future perspectives

Stefano Materia[1]*, Lluís Palma García[1], Chiem van Straaten[2], Sungmin O[3], Antonios Mamalakis[4], Leone Cavicchia[5], Dim Coumou[2,6], Paolo De Luca[1], Marlene Kretschmer[7,8], Markus Donat[1,9]

[1] Barcelona Supercomputing Center, Spain
[2] Vrije Universiteit Amsterdam, Netherlands
[3] Ewha Womans University, Seoul, Republic of Korea
[4] University of Virginia, Department of Environmental Sciences & School of Data Science, Charlottsville, VA, USA
[5] Centro Euro-Mediterraneo sui Cambiamenti Climatici, Bologna, Italy
[6] Royal Netherlands Meteorological Institute
[7] University of Leipzig, Faculty of Physics and Geosciences
[8] University of Reading, UK
[9] Institució Catalana de Recerca i Estudis Avançats (ICREA), Barcelona, Spain

**Abstract**

Scientific and technological advances in numerical modelling have improved the quality of climate predictions over recent decades, but predictive skill remains limited in many aspects. Extreme events such as heat and cold waves, droughts, heavy rain and storms are particularly challenging to predict accurately due to their rarity and non-linear chaotic nature, and because of model limitations. However, recent studies have shown that predictive skill of extremes can be increased when using more sophisticated approaches, indicating that there might be systemic predictability that is not being leveraged.

Recently, numerous studies have been devoted to the exploitation of Artificial Intelligence (AI) to study the predictability and make predictions of weather and climate. AI techniques have shown great potential to improve the prediction of extreme events and uncover their links to large-scale and local drivers. Machine and deep learning, causal discovery, explainable AI, are only some of the approaches that have been tested to both improve our understanding of the processes underlying predictability and enhance prediction skill of extreme events.

Results are promising, especially for hybrid predictions that combine the AI, which can reveal and exploit unknown spatio-temporal connections from data, and climate models, that provide the theoretical foundation and interpretability of the physical world. On the other hand, challenges are multiple in many aspects, from data curation to model uncertainty and generalizability, to the reproducibility of methods and workflows. A few best practices are identified to increase trust in these novel techniques, and future perspectives are envisaged for further scientific development.

# 1. INTRODUCTION

Weather and climate extremes strongly affect many aspects of our society and the environment surrounding us. Being an intrinsic part of a changing climate, there is extensive and well-established evidence that the probability and intensity of extreme events have changed and that these changes will be further amplified in a warming world. Therefore policy-makers and stakeholders are in urgent need of reliable predictions of occurrence probabilities or other aggregated measures of extreme weather on time scales from days to decades ahead. However, the predictive skill of extreme events remains limited, despite the recent improvement in weather and climate prediction systems.

Challenges are multiple:

- The anthropogenic climate forcing has strongly increased since the beginning of the 21st century, mainly due to growing global economy and a decline in the absorbing efficiency of land and ocean $CO_2$ sinks (Canadell et al., 2007). Therefore, studies based on an older attribution period frequently underestimate the effect of global warming on the probability of the unprecedented recent extremes, reflecting the difference between frequencies predicted during the attribution period and frequencies during the out-of-sample verification period (Diffenbaugh, 2020);
- The physical processes that drive or modulate the occurrence of extreme weather events differ among time-scales. Therefore, each times-scale poses unique research questions and requires different models to understand the underlying mechanisms;
- The number of past extreme events covered by reliable and dense observational data is intrinsically small. Many events of the past may be overlooked due to scarcity of the observation availability (Seneviratne et al., 2021): therefore, ensembles of dynamical models are often entrusted with the detection and attribution of their drivers, with possible misinterpretations caused by model limitations.
- Poor representation of key processes and feedback mechanisms between different climate components in climate models, combined with the uncertainties in the initial state, make a complex and chaotic system such as the atmosphere extremely challenging to predict (Faranda et al., 2017).

To face these challenges, the international scientific community has made important steps in the last two decades. On one hand, the World Climate Research Programme (WCRP) has established specific working programs on near-term climate prediction within the new-born Earth System Modeling core project (https://www.wcrp-climate.org/esmo-overview), together with a Lighthouse activity on Explaining and Predicting Earth System Change (EPESC, Findell et al., 2021, see https://www.wcrp-climate.org/epesc). These focus groups aim at developing numerical experiments for subseasonal-to-interdecadal variability and predictability, with an emphasis on improving predictions, and at delivering, through robust process-based detection and attribution, quantitative understanding of the specific changes that are spanning the Earth system.

On the other hand, significant advancements have been made in Earth observation technologies, with enormous improvements in the accuracy and scope of data collected (Guo et al., 2016; Board, 2019; Zhang et al., 2022). The launch of the EU Copernicus programme, the world most ambitious program on Earth Observation, the deployment of new satellite systems (e.g. MODIS, Sentinel) providing high-resolution images of the Earth's surface, the development of new sensors and an increasing collaboration effort between regional space agencies boosted the amount of information available.

This era of "big data" has, in turn, fueled the application of Artificial Intelligence (AI) in many domains of Earth science (Huntingford et al., 2018; Reichstein et al., 2019; Boukabara et al., 2021; Irrgang et al., 2021; Sun et al., 2022). AI here refers to any methodology, including machine learning (ML) and deep learning (DL), in which machines emulate decision-making capacity based on available data. AI algorithms can be used as a set of tools to learn nonlinear relationships between input and output or to extract spatial and temporal patterns from massive datasets, without a priori knowledge of the underlying processes and dynamics of the Earth systems. This makes AI particularly useful for applications for which we do not have a complete theory. For instance, AI can explore subtle or hidden linkages among Earth system's variables, to uncover relevant processes that are not yet implemented in physically-based models. Additional benefits of AI include its flexibility to employ a wider range of input variables, such as novel remote sensing observations, as opposed to physics-based models which only consider input variables traditionally assumed to be correlated with a target variable. In this way, AI can assist us in exploiting the full potential of big data, leading to new insights into Earth system processes that can inform model development and evaluation. This is also the case for climate science, including extreme climate predictions.

Progress in AI-based forecasting on weather time-scales, i.e. less than 10 days, has been remarkable in the last few years. In parallel with the rapid rise of AI, forecasting institutes worldwide and Big Tech companies have seized upon the opportunity to improve weather forecasts, gaining skill comparable to that of state-of-the-art dynamical predictions (Pathak et al., 2022; Bi et al., 2023, Lam et al., 2023). Compared to the short time-scales, progress on the subseasonal to decadal (S2D) time-scale has been less striking. A fundamental challenge facing climate scientists is the limited amount of independent training data, roughly one or two orders of magnitude smaller than for weather time-scales. This has long hampered the development of long-lead forecasts of extremes like drought and warm spells, which likely have at least some predictability at the S2D scale. In truth, it has been suggested that the predictability of the climate system beyond the deterministic time-scale is much larger than that implied by the CMIP6 generation of climate models taken at face value (Scaife and Smith, 2018; Smith et al., 2020). In fact, an increasing number of articles has been published since the "S2S reboot" opinion paper (Cohen et al., 2019), that claimed that ML techniques mostly developed in computer science could be adopted by climate forecasters to increase the accuracy of predictions at subseasonal to seasonal (S2S) scale.

So far, the development of AI methods for the prediction of extreme events has been overlooked, while this topic is critical for applications and assessment of their usefulness in real life (Watson, 2022). This review explores the potential of AI to improve the prediction of extremes, and to reveal their links to large-scale and local drivers. Through an excursus on recent literature about recent application of AI for climate extreme predictions, prospects brought by the combination of statistical and dynamical methods and the challenges of the data-driven approach are discussed, together with future perspectives.

## 2. AI-BASED PREDICTION OF EXTREME EVENTS AT SEASONAL TO DECADAL TIME-SCALES

### 2.1 Predictions beyond the weather time-scale

The development of weather and climate extremes depends on a favourable initial state, the presence of large-scale drivers, and positive feedbacks, as well as stochastic processes (Sillman et al., 2017). Time-scale is what marks the distinction between two types of predictions: while specific weather situations can be predicted for up to 10-15 days in advance, weather as such becomes deterministically unpredictable beyond this time scale (Lorenz 1969, 1982). Climate predictions target longer forecast times, e.g. seasons, years or decades ahead. Such climate predictions are necessarily probabilistic, and predict tendencies or the climate system rather than individual events (Meehl et al 2021).

Few days ahead of the occurrence of an extreme event, it is possible to make predictions of its occurrence and amplitude, with considerable detail about its location, onset and duration, if the local and remote processes leading to its generation are properly initialised and well predicted (Domeisen et al., 2023). Deterministic forecasts can be made up to ten days ahead for specific extremes mostly linked with long-lasting atmospheric patterns (e.g. heatwaves, Fragkoulidis et al., 2018). Beyond the predictability limit, it is possible to forecast a probability distribution of the intensity and duration of an extreme (Domeisen et al., 2022), a temporal propensity for the occurrence of extremes (Prodhomme et al., 2022), or a change in their frequency (Cai et al., 2014). These characteristics are those potentially captured by a climate prediction (Fig. 1), which is the focus of this review, while deterministic weather forecasts will not be discussed since the scientific questions and the approaches to fulfil them can differ much.

On (multi-)weekly to decadal time-scales, both local and remote physical processes can contribute to the predictability of extreme events (Fig. 1). These mechanisms take action seamlessly across time-scales, but their relative contribution varies across the range of forecast times and relates to the degree and time-scale of interaction between the troposphere and the slower-evolving climate components. In general, land-atmosphere coupled processes convey predictability over time-scales between weeks and a few months (Miralles et al., 2019; Materia et al., 2022), while stratospheric variability and stratosphere-troposphere coupling have been found to provide potential predictability from subseasonal to multi-annual time-scales (Kidston et al. 2015, Scaife et al. 2022). Ocean-atmosphere coupled mechanisms both in the tropics and the extra-tropics act on a wide range of time-scales from weeks up to multiple years (Santoso et al., 2017; Ossó, et al., 2020, Ding et al., 2022), while the effects of very slow modes such as the Atlantic Multi-decadal Variability span for decades (Zhang et al., 2019b). These variations and their interlinks with the troposphere eventually act as boundary conditions for the atmospheric circulation (Shukla et al., 1998), contributing to generate oscillations (MJO, e.g. Zhang 2005, Zhang et al., 2020, or ENSO, e.g. Rasmusson and Wallace, 1983, Capotondi et al., 2015) and patterns (blocking, quasi-stationary waves, e.g. Reinhold and Pierrehumbert, 1982) able to give the atmosphere persistence characteristics.

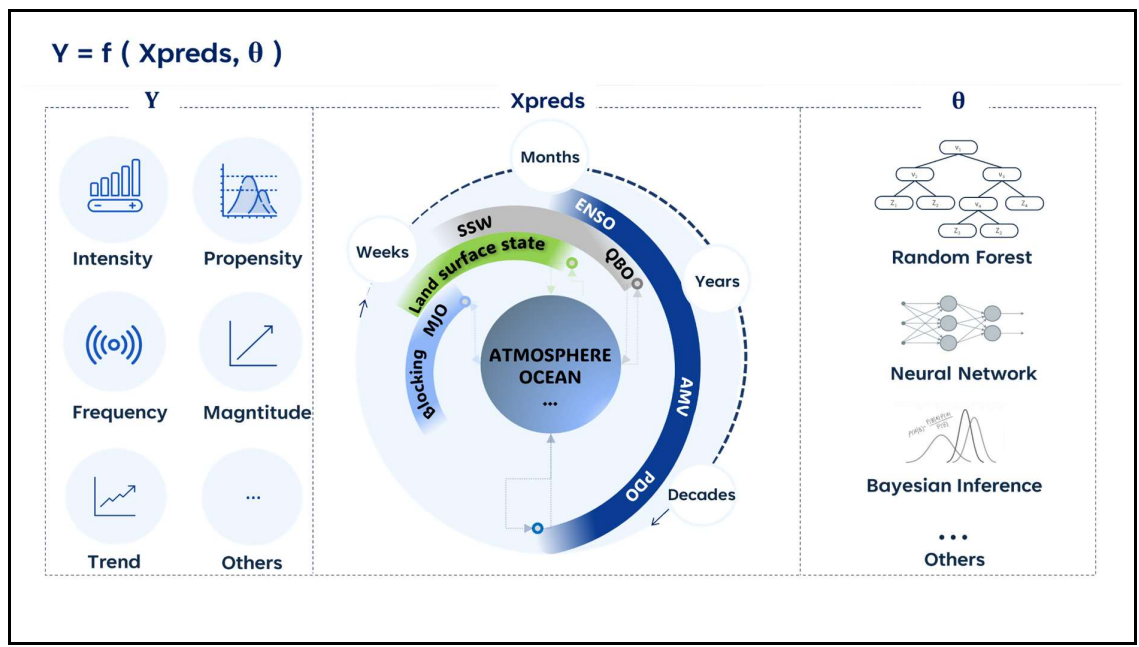

Fig. 1. Schematic representation of the AI-based climate prediction function. **Y** represents the **predictand** or **target**, that is a measure of certain aspect characterising the extremes of interest (e.g. intensity, frequency, etc.). **Xpreds** are the **predictors**, such as modes of variability, or variables describing the state of an Earth System component that affects the Earth system component where the extreme takes place (illustrated as the circle at the center of the figure). $\theta$ represents the **parameters** of the AI algorithm used to train the model (*f*).

These modes of variability and couplings between Earth system components act as drivers for the insurgence of climate extremes. Drivers acting at different time-scales may interfere with each other and affect the amplitude and characteristics of potential extremes. For example, a negative phase of the PDO is associated with prolonged wet phases in southeastern US winter (Fuentes-Franco et al., 2016). At the seasonal time-scale, the expected damp anomalies are often reduced during El Niño years, while they are reinforced in winters marked by La Niña (Wang et al., 2014). However, during a phase of an active MJO, the extratropical response can amplify or mask the interannual ENSO signal for a few weeks in southeastern US, potentially resulting in wet or dry extremes of the opposite sign than that anticipated by the ENSO phase (Arcodia et al., 2020).

Aside from the skill obtained via initialisation with observed climate conditions, allowing coupled climate models to capture mechanisms of internally generated climate variations, the large trends in boundary conditions, such as greenhouse gases and aerosols, may also represent a significant source of predictability (Meehl et al., 2009). This effect becomes increasingly important with lead time, and the large impact of the background warming has often been shown in decadal predictions (Bellucci et al., 2015; Smith et al., 2021). However, the contribution of trends to predictability may be also detected at subseasonal and seasonal time-scales (Butler et al., 2021, Patterson et al., 2022, Wulff et al., 2022), particularly in the summer when the year-to-year variability is lower, potentially increasing the skill of extremes forecasts (Prodhomme et al., 2022).

The potential of AI to enhance the predictive skill of climate extremes across various time-scales holds significant promise for advancing our understanding and preparedness on a societal level (Huntingford et al., 2019). The development of sophisticated data-driven approaches offers the opportunity to harness the interconnections of climate drivers and processes across scales, including local and remote interactions, stratosphere-troposphere couplings, and ocean-atmosphere interactions (see Fig. 1). AI algorithms can effectively analyse vast amounts of climate data and identify complex patterns that influence extreme events in a nonlinear way, as determined by the data itself, without the use of potentially biased numerical Earth system models. Moreover, methods of explainable AI and causal discovery algorithms (e.g. Lundberg and Lee, 2017) can help us better understand the relative contribution of various climate drivers, such as PDO, ENSO, and the AMV, and how they may interact to modulate climate extremes.

Given the large appeal to acronyms that describe both climate modes of variability and AI algorithms, we summarize and define those recurrently used in this article in table 1.

|  | **Acronym** | **Extended name** |
| --- | --- | --- |
| **Climate** | **AMV** | Atlantic Multidecadal Variability |
|  | **ENSO** | El Niño Southern Oscillation |
|  | **MJO** | Madden Julian Oscillation |
|  | **PDO** | Pacific Decadal Oscillation |
|  | **QBO** | Quasi-Biennial Oscillation |
|  | **S2D** | Subseasonal to Decadal |
|  | **S2S** | Subseasonal to Seasonal |
|  | **SPI** | Standardized Precipitation Index |
|  | **SSW** | Sudden Stratospheric Warming |
| **Algorithms** | **ANN** | Artificial Neural Network |
|  | **CNN** | Convolutional Neural Network |
|  | **LSTM** | Long Short-Term Memory |
|  | **RF** | Random Forest |
|  | **XAI** | eXplainable Artificial Intelligence |
|  | **XGBoost** | eXtreme Gradient Boosting |

Table 1. List of acronyms often used in this article

### 2.2 Extreme temperatures

Hot extremes are exacerbated by anthropogenic climate change and they often manifest in the form of heatwaves (Thompson et al. 2023; Barriopedro et al. 2023). In a warming world, cold spells are expected to become less frequent, durable and intense, but still can pose significant challenges to human activities especially in the northern mid-latitudes during boreal winter (Matthias and Kretschmer, 2020; Tomassini et al., 2012).

Classifying temperature extremes meaningfully can be challenging, since definitions differ according to the specific scientific questions, sectoral application and time-scale. Temperature extremes are often detected as (consecutive) days which exceed a certain threshold (e.g. Perkins and Alexander, 2013; Sillmann et al. 2013; Russo et al., 2014), but different approaches, e.g. based on cumulative metrics, are also used (Russo & Domeisen 2023). While current extreme indices cover a wide range of extreme temperature attributes, such as amplitude, duration and frequency (Zhang et al., 2011), the variety of definitions is much broader when applied to predictions, in an attempt to suit the challenges of such a difficult task. Prodhomme et al. (2022) introduced the concept of *heat wave propensity*, namely the tendency of a season of being inclined to heat waves, claiming that a seasonal forecast may more easily predict such a characteristic. Ragone et al. (2018) used a definition that merges temperature and geopotential height anomalies to detect long-lasting heat waves as temporal and spatial averages, improving the statistics of extremely rare events using rare event algorithms (Ragone and Bouchet, 2021).

The ability of AI to forecast various aspects of temperature extremes on S2D time-scale is demonstrated in several recent studies and a variety of methods have been used. Decision-tree-based ensemble methods like Random Forests (RF) and XGBoost (He et al 2021; van Straaten et al. 2022; Weirich-Benet et al. 2023; Kiefer et al. 2023) are often chosen for their robustness against overfitting. Neural networks (NN) like CNNs, LSTMs, and transformers have the potential benefit of directly taking in spatio-temporal information (Chattopadhyay et al. 2020; Lopez-Gomez et al. 2022; Jacques Dumas et al. 2022; Miloshevich et al. 2023). Being the most data-hungry, and given the limited occurrence of extreme temperature events in historical data, this lack is often circumvented by training NNs on data from numerical climate model simulations (Chattopadhyay et al. 2020 Jacques Dumas et al. 2022; Miloshevich et al. 2023).

Successful applications of multiple linear regression approaches, accompanied by tailored filtering procedures to extract only the predictors representing the predictable drivers also exist (Miller and Wang 2019; Pyrina et al. 2021; Trenary and DelSole 2023). This links to an important goal besides the ability to produce skillful forecasts, namely the use of AI to identify and disentangle drivers, and thereby supplement incomplete theory. For example Polkova et al. (2021) used causal discovery algorithms to understand and predict marine cold air outbreaks in the observations and a seasonal prediction model. Suarez-Gutierrez et al. (2020) uses Multiple Linear Regression to understand dynamical contributions to European heat extremes. By combining RFs with explainability tools, van Straaten et al. (2022) investigate the influence of atmospheric, oceanic and land surface states on different time-scales on the occurrence of heat extremes. The same idea is applied by Kiefer et al. (2023) for cold extremes. In both cases, AI algorithms are applied to learn the physically known relationships in the data.

## 2.3 Droughts

Drought is a complex natural hazard which is difficult to define due to multiple causing mechanisms involved at different spatial and temporal scales in its occurrence (Wilhite, 2000). Accordingly, various drought indices based on individual or multiple hydro-climatic factors have been developed to detect and predict drought events. Standardised precipitation index (SPI: McKee et al., 1993) or Palmer drought severity index (PDSI: Palmer, 1965) were proposed for meteorological droughts, Streamflow Drought Index (SDI: Nalbantis and Tsakiris, 2009) for hydrologic droughts, soil moisture and vegetation percentiles (Yihdego et al., 2019) or the SPI including the effect of evapotranspiration (SPEI, Vicente-Serrano et al., 2010) for agricultural droughts. The most widely used AI algorithms for drought predictions include XGBoost (Zhang et al., 2019a; Gibson et al., 2021), RF (Park et al., 2018; Gibson et al., 2021), support vector machine (Ganguli and Reddy, 2014; Liu et al., 2017, Mokhtarzad et al., 2017) and DL algorithms such as Multi-Layer Perceptrons, CNN, or LSTM (Poornima and Pushpalatha, 2019, Sahoo et al., 2019, Adikari et al., 2021; Dikshit et al., 2021, Gibson et al., 2021; Mokhtar et al. 2021). These AI-based models seek to predict drought indices over different forecast times and show more robust and accurate performance than commonly used traditional statistical models (e.g. regression models and autoregressive moving average models) as they can effectively handle large amounts of non-linear data (Mokhtarzad et al., 2017; Poornima and Pushpalatha, 2019). AI-based models show satisfactory performance also in estimating drought properties such as drought severity level or propagation probability (Al Kafy et al., 2023, Jiang et al., 2023).

More recent studies have drifted towards combining AI algorithms (e.g. Ahmed et al., 2022; Danandeh Mehr et al., 2022; Xu et al., 2022). For instance, Ahmed et al. (2022) proposed one-dimensional CNN combined with a gated recurrent unit for evapotranspiration forecasting, so the model can better capture the time series dependence. Long short-term memory (LSTM) is one of the widely-used algorithms in this respect; it has gained popularity in previous hydrometeorological prediction studies due to the ability to learn long-term dependencies across time steps in sequential data. For instance, Danandeh Mehr et al. (2022) suggested a CNN-LSTM model for drought forecasting one month ahead, and showed its better performance against counterpart benchmark models.

While more comprehensive reviews on the application of AI in drought prediction can be found from, e.g. AghaKouchak et al. (2022) or Prodhan et al., (2022), most of the existing studies are tightly geographically focused. Therefore, model performance is highly dependent on the specific study conditions, such as study region, drought indicator, or input variables considered, making it difficult to generalise the main findings from one study to another.

### 2.4 Cyclones and heavy precipitation

Synoptic scale cyclones, both in the tropics and the mid-latitudes, are among the extreme events causing the largest economic damage (Mendelsohn et al 2012) due to the associated heavy precipitation, strong winds and storm surges. There is indication that climate change could exacerbate the severity of such extremes (even if not necessarily the frequency of their occurrence) (Knutson et al. 2020). However, predicting their variability on S2D time-scales remains challenging (Befort et al 2022). Heavy precipitation events are not always associated with large-scale weather systems such as cyclones or fronts: many impactful events are instead linked to short-lived, small-scale severe convective events. These extremes are even more challenging for the realistic representation in operational climate prediction systems, since their spatial resolution is too coarse to allow the explicit representation of convection. Indeed, numerical climate prediction systems skill for extreme precipitation has been shown to substantially decrease beyond a few days in most of the regions where it has been analyzed (e.g. King et al 2020, Rivoir et al 2022).

AI techniques have been successfully applied to improve the prediction of both cyclones and heavy precipitation events from a number of different angles. A particularly promising framework is given by hybrid statistical dynamical predictions (see also Section 3). In this setting, the goal is to improve the skill of numerical prediction systems (e.g seasonal predictions) in representing weather extremes by finding relationships between the large-scale drivers (on which the dynamical model has good skill) and the extreme events. This approach has been applied targeting either large-scale extreme events (e.g. tropical or extratropical cyclones) or directly the precipitation field (Scheuer et al., 2020; Specq and Batté, 2020). Examples of ML/DL methods used to predict tropical cyclone occurrence include CNNs (Fu et al 2022), RFs (Tan et al 2018). Both approaches have advantages and disadvantages. RF provides more interpretability on the role of different drivers, while CNN approach appears to be more powerful in terms of prediction skill, even though the increased skill is associated with a decrease in interpretability. de Burgh-Day & Leeuwenburg (2023) proposed to systematically perform ablation studies on the model as a possible way to overcome the interpretability issues of DL models while retaining their good skill. A similar RF-based approach has also been applied for subseasonal predictions of precipitation (Zhang et al. 2023) showing promising results.

Several AI applications also contribute to the improved prediction of climate extremes not by performing directly the prediction task themselves, but enhancing the way climate forecast models output is processed. Following the classification proposed in Section 3, those techniques belong to the second group (serial approach). Applications aimed at improving the representation of wind and precipitation fields deserve a particular mention. They exploit deep learning algorithms trained on high-resolution observations to improve the representation of precipitation (Vosper et al 2023) or wind patterns (Yang et al 2022) associated with cyclones. They can therefore be considered a form of downscaling.

## 3. HYBRID CLIMATE PREDICTIONS

Hybrid climate predictions combine numerical forecasting techniques with AI methods. This has emerged as a promising approach for improving the accuracy and reliability of climate predictions as it ideally takes advantage of the strengths of both techniques: the physical consistency of dynamical models, and the flexibility and adaptability of ML/DL models. In this section, we will review some of the recent advances in hybrid modelling for predicting weather and climate extremes. We follow the typology developed for hydrological forecasting by Slater et al (2023) and focus on three main areas: 1) a fully coupled approach in which AI improves the parametrization and initialization of climate models,

2) a serial approach in which climate model outputs are post-processed with ML, or in which multiple model outputs are blended with ML and, 3) a statistical-dynamical approach in which climate models are used to train a data-driven model.

The first main area addresses the major challenges in climate modeling to accurately represent small-scale processes, such as cloud formation and convective storms, which are critical for predicting some types of weather and climate extremes. AI can help improve the representation of these processes in climate models. While this type of hybrid modelling does not explicitly target the predictions of extreme events in the first place, prediction of extremes may benefit from the improved simulations of relevant mechanisms. For instance Rasp et al. (2018) employ a fully connected NN to parametrize convection and cloud processes within the atmospheric column of a climate model, similarly to Gentine et al. (2018). In these cases the NN is first trained in multi-scale cloud-resolving simulations in which the network learns to emulate fine-scale modelled processes, such that it can fulfil the same role when coarser-scale forecasts are made at the time of inference (Figure 2A). Steps towards coupling are being made for land-surface parametrizations as well (ElGhawi et al. 2023). A common prerequisite for good coupled performance is the stability of the AI algorithm and an ability to generalize to unseen situations, which is why these algorithms are usually designed to adhere to known physical relations (see Section 5).

The second area of hybrid prediction uses machine learning techniques to post-process climate model outputs. This can be used to bias-correct and/or downscale model outputs that suffer from inaccurate representations. A statistical model is tasked to learn systematic biases between forecasts and observations, which for probabilistic forecasts can concern a correction of the full probability distribution, such that also the occurrence probability of extremes is reliably estimated. Errors in model-generated heavy precipitation are for instance corrected by learning the precipitation patterns generally induced by ENSO and applying those to forecasted ENSO (Doss-Golin et al. 2018, Strazzo et al. 2019), which is a perfect-prognosis approach (Fig 2B). Another way is that systematic errors between forecasts and observations are learned directly with a Model Output Statistics algorithm (Fig 2C), of which simple ones have been applied early since the development of weather models (Glahn and Lowry, 1972). Current ML techniques are an improvement because they can learn non-linear relations (Vannitsem et al. 2021, Haupt et al. 2021, Schulz and Lerch 2022) and thus apply corrections that depend on the weather conditions in which an extreme is occurring. On subseasonal time scales studies have employed fully connected NNs (Fan et al. 2021, van Straaten et al 2023), convolutional and UNet-type NNs (Scheuerer et al. 2020, Horat et al. 2023), RFs (Zhang et al. 2023), an ensemble of regression models (Hwang et al. 2019), or Bayesian methods (Schepen et al. 2014, Strazzo et al. 2019, Specq and Batté 2020). In these studies the post-processing is applied to produce probabilistic forecasts of weekly, bi-weekly or monthly accumulated precipitation or average temperature.

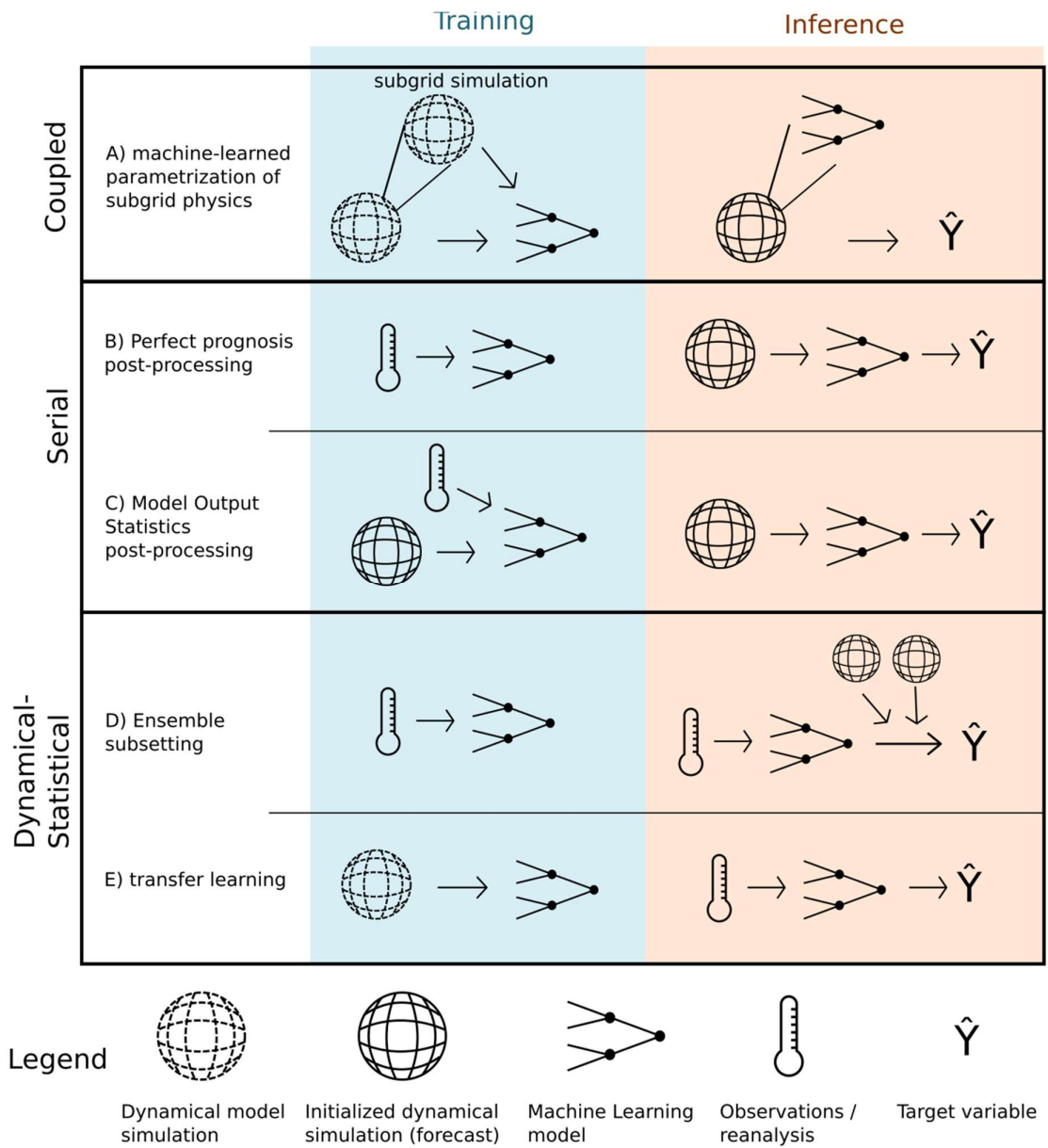

Fig 2. Schematic representation of methods to combine dynamical models with machine or deep learning to create hybrid predictions.

Besides improving forecast skill, statistical post-processing models can also serve to evaluate the dynamical models that they correct. To correct weather dependent systematic biases the ML method will use predictors related to the weather-type and its preceding sources of predictability. This way post-processing relates error characteristics to the circumstances under which they occur, and can thus expand upon the physical understanding of extremes, which would be limited when using numerical models only. Examples are Silva et al. (2022), Mouatadid et al. (2023), van Straaten et al. (2023)

Finally, a third broad category of hybrid forecasting can be called statistical-dynamical (Slater et al. 2023). Here the idea is to combine the prediction from a dynamical model, with that of an empirical, purely data driven approach, where the data-approaches as well as the way to combine prediction vary (Fig. 2D). One way is to use statistical methods to provide first-guess prediction of an important state variable (e.g. NAO), related to the extremes of interest, and then weigh or select dynamical simulations accordingly (Dobrynin et al. 2018, Neddermann et al. 2019, Polkova et al. 2021). In this context, the first-guess prediction has been made based on expert-informed regression models (Dobrynin et al. 2018), as well as based on using causal discovery algorithms (Polkova et al 2021).

Another example is the training of an ML model on climate model simulations, integrating data from large climate model ensembles. This methodology, known as transfer learning (fig. 2E, Weiss et al. 2016) represents a promising approach towards expanding the size of available training sets (Ham et al., 2019; Andersson et al., 2021; Gibson et al., 2021). The long records of simulated data produced by climate models are an alternative to the observational datasets employed in classical empirical approaches. Such records can cover hundreds of years with simulated climate conditions under various forcing scenarios incorporating different greenhouse projections. This information allows ML models to assimilate a more statistically robust version of the climate system and its future trends, opening the door to a better generalization/extrapolation under future climate conditions. In addition, multiple models or ensemble realizations can be combined to expand the dataset by some orders of magnitude, particularly with models that run large ensembles. Similarly, data-driven models based on causal discovery algorithms can be fitted on dynamical simulations not to eventually make predictions, but to directly evaluate the presence of known links in the climate system (Di Capua et al. 2023).

## 4. CHALLENGES IN THE ARTIFICIAL INTELLIGENCE APPROACH

The vehemence with which AI has irrupted in the climate prediction discussion comes at the cost of a large number of unresolved challenges, which need to be addressed in order to build trust in this emerging technology at the service of climate applications. We identify five major areas of challenges and we propose a few best practices that should be acknowledged and carried out in future studies.

### 4.1 Data and processing

Extreme events are rare by definition. In addition, unprecedented events are becoming more likely in a non-stationary climate (White et al., 2021), and this poses an important scientific challenge to the opportunity of improving climate predictions of extremes using AI. In fact, the scarcity of historical data, and the absence of such utmost events in the past, make any statistical approach primarily based on observations predestined to a failure (Miloshevich et al., 2023). The physical processes involved in the development of climate extremes could have time cycles ranging from weeks to years, different seasons have different predictive relations, and many climate variables are temporally correlated at multiple time scales (He et al., 2021), therefore it is difficult to obtain sufficient (effective) samples to learn from. This problem becomes more evident with increasing length of forecast periods, which strongly limits the possibility to verify a sufficient number of time-independent forecasts: multi-annual (five to ten years) predictions, whose training sample meant to rely on current atmospheric reanalysis (e.g. ERA5, Hersbach et al., 2020), would not have more than a dozen samples to learn from.

As discussed in section 3, transfer learning has the potential to enlarge the learning sample using climate model data. This implies a realistic representation of relevant processes and a characterization of model systematic errors through process-based studies (Eyring et al., 2019). Thus, the climate models selection becomes a crucial part of the pipeline with significant implications for the final performance of the ML model. Recent studies have followed various approaches to ameliorate this issue. Some choose a single model known to correctly represent the physical processes involved in the targeted tasks (Gibson et al., 2021; Miloshevich et al., 2023). Others pool many models into the training set, allowing the ML algorithm to capture the robust signal within all the simulations (Ham et al., 2019; Ling et al., 2022; Pan et al., 2022). Furthermore, this procedure can be extended by a second training loop, known as fine-tuning, where the ML model is further trained using available observations (Ham et al., 2019; Andersson et al., 2021; Pan et al., 2022), thus accounting for the biases present in the first training loop.

Despite having alternatives to extend the training dataset through different approaches, the so-called imbalanced learning problem (He & Ma, 2013) is inherent to extreme forecasting. Imbalance learning refers to the fact that, by definition, extremes (events in the distribution's tails) will always be less frequent than events closer to the average of a distribution. The more extreme the events targeted are, the larger the ratio between extreme and non-extreme samples will be. Imbalanced learning leads to less confident models predicting extreme states over normal ones, leading to unskilful predictions if not adequately addressed. So far, the proposed solutions have often employed resampling techniques where either the minority class goes through oversampling or the majority class goes through undersampling: this could be done through randomly sampling the current available samples (Miloshevich et al. 2023), or by generating new samples using interpolation (Synthetic Minority Oversampling Technique, SMOTE; Chawla et al., 2002). Still, it is essential to remark that none of the re-sampling methods will add extra information out of the fewer samples of extremes available, and some studies point out that unreliable probabilities may characterise ML models trained using re-sampling techniques (Fissler et al., 2023).

Several types of extreme events are clearly affected by the background global warming trend. An increase in heat waves frequency and magnitude (and the concurrent decrease of cold spells) is undoubtedly occurring virtually everywhere in the world, while heavy precipitation, storm intensity and drought frequency/duration locally show increments that are likely to be associated with a warming climate. This poses a question on how to deal with this anthropogenic constraint, to isolate its induced predictability from the natural predictability of the system. Many studies simply perform an out-of-sample pre-processing: as the trend is global and deterministic, it is easy to remove by detrending the training time-series before the application of the learning algorithm (e.g Weirich Benet et al., 2023, ). While this approach allows to target only the natural drivers of the studied extremes, they possibly exclude an important source of predictability across multiple time-scales (Bellucci et al., 2015, Prodhomme et al., 2022, van Straaten et al., 2022, see also Section 2.1). There is no general solution to this issue, and the treatment of trends mostly depends on the scope of the works AI-based predictions are designed for, or questions being targeted. Improving prediction skill for e.g. climate services would benefit from the employment of the trend in the learning dataset, while removing it is sensible if the aim is uncover potential drivers for the studied extreme, and separate the human-made contribution from the natural variability (Zeder and Fischer, 2023). The latter approach would also require an additional choice on the specific method of trend removal (Frankcome et al., 2015), on what variables, and how to deal with variables indirectly affected by the background trend (e.g. soil moisture), whose variability is dominated by other factors. Efforts in this sense have been hardly carried out in the framework of ML for climate predictions of extremes.

### 4.2 Uncertainties

Due to the inherent complexity and chaoticity of the climate system, the relationship between predictors and predictands is intrinsically probabilistic; there is no one-to-one correspondence between the two. Thus, climate predictions of extreme events are subject to significant uncertainties. As a result, the magnitude of those uncertainties (i.e. aleatoric uncertainties) plays an important role in determining the potential predictability of an extreme event (Lucente et al., 2022). Likewise, uncertainties originating in the data and the ML method (i.e. epistemic uncertainties) also affect each model prediction. Hence, probabilistic models are the most suitable to express the combined uncertainties (van Straaten et al.

2022; Miloshevich et al. 2023). The resulting distribution of possible future states better characterises the nature of the prediction problem. Such distributions are most informative to decision-makers when they are 'calibrated', meaning that issued probabilities are consistent with observed occurrences (Gneiting et al. 2007). Many ways exist to produce probabilistic forecasts with ML, but not all output explicit uncertainty distribution (Luo et al., 2022). Thus, proper and reliable probabilistic ML modelling remains a vital research line in climate prediction of extremes.

Regarding probabilistic approaches, one method specific to ANN makes use of the dropout Monte Carlo approach. In this setting, during the training of the network, a small fraction of neurons are randomly "dropped out", (i.e., deactivated) in each iteration of the optimization process. The end result of dropout training is that the trained network is more robust (Scher and Messori, 2021). Also, during inference (after training is completed), one can quantify uncertainty by sampling many different predictions, each time deactivating a different set of random neurons. Other methods, applicable to any machine learning model, consider training an ensemble of models on subsets of the data or by starting from different random seeds (e.g. Weyn et al. 2021), or else by retaining, instead of the best estimator, a number of "best estimators" to optimise the hyperparameters of a machine-learning model. In all these cases uncertainty is treated in a post-hoc fashion, meaning that these methods usually do not guarantee calibration.

### 4.3 Interpretability and causability

One of the most important pitfalls of using AI for predicting climate extremes is that the model's decision-making is in many cases unknown. With the exception of only a few methods (e.g., see linear/logistic regression, decision trees and newly introduced DL; see Chen et al., 2019; Agarwal et al., 2021; Barnes et al., 2022), the majority of ML/DL algorithms are highly complex and non-transparent, which makes their decision making non-interpretable to the user. Although for some applications high performance might be sufficient and interpretability might be less of an issue (e.g., machine translation or text generation), for climate applications and extreme event prediction, interpretability is fundamental because of the potentially devastating effects a false prediction might have on society. To address the issue of interpretability, the computer science community has developed tools that can be used to explain the predictions of black-box AI models in a post prediction setting (Buhrmester et al., 2019; Tjoa and Guan 2019; Das and Rad 2020); the so-called eXplainable AI (XAI) tools. XAI has recently attracted much attention in a number of fields, including the geosciences (e.g., McGovern et al. 2019; Ebert-Uphoff and Hilburn 2020; Toms et al. 2020; Barnes et al. 2020), where many applications show its potential to make the black box more transparent, build model trust, fine-tune poor performing models and lead to new scientific insights (McGovern et al., 2019; Mamalakis et al., 2022a).

Despite XAI showing great potential, there are still challenges. First, XAI tools are imperfect themselves, and the degree to which they are a true representation of the AI model may depend on the application and the prediction setting (Mamalakis et al., 2022b). Some studies have highlighted that XAI methods may face issues with their faithfulness to the ML/DL model, the comprehensibility of their results, and reproducibility (Mamalakis et al., 2022c; 2023). Due to the above potential pitfalls of XAI tools, there is some work that argues against using XAI and instead in favour of developing inherently interpretable AI models (Rudin, 2019; Rudin et al., 2022). Lastly, even if one assumes that XAI tools are faithful to the network, their insights should only be used to highlight sources of predictability and not to infer causality. The reason for the latter is that unless properly designed, an AI model might be using relationships that are not physical (i.e., spurious correlations) to make its predictions. The issue of not being able to draw causal conclusions from XAI applications refers to the limited "causability" of these tools (Holzinger et al., 2021; Mamalakis et al., 2022b) and is an important remaining challenge. Physics-guided AI (also known as knowledge-guided or physics-informed) constitutes one of the ways some researchers try to impose physical realism in the prediction algorithm and limit the effect of spurious correlations on the training (see section 5), but this area of research is still in its infancy.

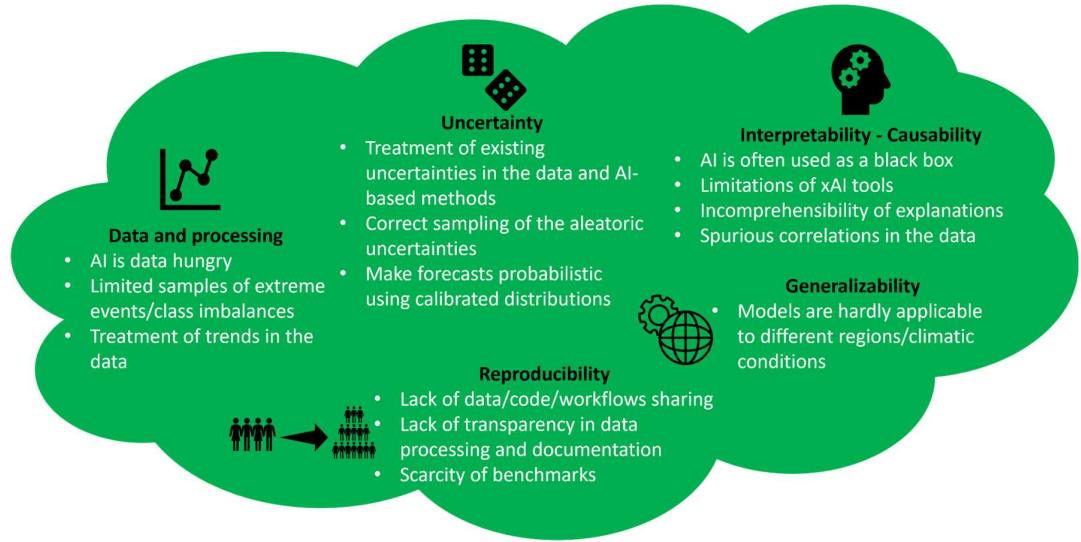

**Figure 3.** The current challenges facing the prediction of climate extremes with AI. Major areas of difficulty (Data and processing, Uncertainty, Interpretability and Causability, Generalizability, Reproducibility) are shown in black, and accompanied by a complementary icon, while related tasks discussed in the text (see section 4) are shown in white.

**4.4 Generalizability**

Generalisation refers to the model's capacity to make accurate predictions beyond the spatio-temporal boundaries of the training datasets. Traditional ML/DL algorithms rely on the assumption that training and testing (unseen) data are identically distributed and that the relationships between input and target variables learned during model training are equally valid on the testing data. However, in climate science applications, this assumption often fails to hold when deploying models to predict extreme values that lie outside of the climatological distribution of the data used for training. This issue of out-of-distribution generalisation can lead to considerable degradation in model performance, particularly under a warming climate that shifts spatial and temporal distributions of Earth system variables, as the current relationships that best describe predictor variables and climate extremes may no longer be valid in the future (O'Gorman and Dwyer, 2018, Rasp et al., 2018, D'Amour et al. 2020). Similarly, predicting extreme events across diverse climatic regimes also poses a generalisation problem. The accuracy of ML/DL models can vary substantially when making predictions on contrasting climatic conditions, for instance, when a model is trained with data obtained from humid climatic regions and then used in arid regions (O et al. 2020; Meyer and Pebesma, 2021). Therefore, understanding model performance on previously unseen conditions without access to labelled data is one of the fundamental challenges associated with enhancing the robustness of ML/DL models. Recent studies have shown that a large, diverse training data from a wide range of climatic regimes is crucial for achieving robust model performance even when a model is used over a limited geographic region; AI can infer past or future temporal variabilities of extreme events from contemporary spatial variabilities (space-for-time approach) (O et al., 2020; Wi and Steinschneider, 2022). Physics-informed ML/DL also shows a promising step forward for enhancing model robustness (Wi and Steinschneider, 2022); we note, however, that the out-of-distribution generalisation has not yet been sufficiently explored in the context of climate extreme predictions.

### 4.5 Reproducibility

Reproducibility refers to the ability of researchers to independently replicate and verify the results of a study, given access to data, methods, and procedures employed to obtain them. Reproducibility is a minimal prerequisite for the creation of new knowledge and scientific progress because it ensures that findings are reliable and trustworthy, particularly when the approach used is innovative and highly transformative. This is also the case for ML applications in climate predictions, where incredibly fast development has occurred in a matter of a few years, with new techniques, algorithms and workflows published or simply posted on non-peer reviewed online platforms at an accelerating rate.

However, more than 60% of earth scientists stated that they failed to reproduce other researchers' work, while more than 40% admitted they could not reproduce their own experiment (Baker, 2016), raising concerns about a "reproducibility crisis" which is amplified in AI literature (Hutson, 2018). The most basic problem is that researchers often do not share their source code, due to a variety of reasons: the code might be a work in progress, or jealously held by a researcher eager to stay ahead of the competition or that simply does not feel comfortable with their own scripting skills (Gundersen and Kjensmo, 2018). In many cases, not even the datasets used for training are made available to the community.

While sharing codes and data is a crucial point, a precise knowledge of what exactly was investigated and how the experiments were conducted is imprescindible: the more details the documentation contains, the easier it is for independent researchers to reproduce the results, the larger is trust in the results. An exhaustive documentation also reduces the effort required to make the experiment, therefore lowers the barriers for others to actually run it. In drafting this review, a number of published papers were discarded from the references because the description of methodology was poor or inaccurate, making the findings questionable.

Finally, the scarcity of common datasets and evaluation metrics make intercomparison between climate extreme prediction studies difficult. So-called benchmarks can make algorithms quantitatively inter-comparable and boost competition in such an emerging research line. If well-curated, benchmark datasets may help people with different expertise (i.e. climate and computer science) work on a common problem (Rasp et al., 2020). However, designing standardised datasets can be extremely complicated, as climate is a high-dimensional and multi-faceted problem where each question can be very specific (Dueben et al., 2022). Efforts to build benchmark datasets are currently ongoing for weather-scale forecast times, at which the atmosphere is still deterministic (WeatherBENCH, Rasp et al., 2020, 2023), or multi-decadal time-scale, where the response of the climate system is largely driven by the socioeconomic scenario (ClimateBENCH, Watson-Parris et al., 2022). However, no such benchmarks are yet available for climate predictions, where there is a large range of time-scales at which processes occur, with many interactions taking place within and across scales.

---

*Sidebar*
**Best practices to improve trust in AI-based forecast of extremes**

While artificial intelligence provides tools to target potential windows of predictability and eventually improve prediction skill of extreme events, this skill by itself is insufficient. Trust is essential for any early action, which is the ultimate goal of forecasts. Currently, AI-based forecasting generally suffers from a lack of trust for multiple reasons: (1) methods are often used as black-boxes with the sources of predictability unexplained, (2) there are many technical pitfalls that have resulted in exaggerated claims on ML-based skill, and (3) the exact data processing is often non-transparent and non-reproducible.

To overcome this lack of trust, we recommend the following 'good practices':

1. An effort in understanding sources of predictability and underlying physical mechanisms is required. Interpreting machine learning models should be a top priority, with interpretability focusing on causality instead of association. Explainable AI can provide insights into the sources of predictability, but a commitment towards interpretable models is highly encouraged (Rudin, 2019).

2. Proper and suitable quantification of uncertainties should be prioritized in order to minimize epistemic uncertainties and sample all possible aleatoric uncertainties.

3. Validation should be described step by step, and preferably multiple cross-validation approaches should be tested (Sweet et al. 2023), being aware of the possibility of information leakage from train to test data (Risbey et al., 2019). Ideally all pre-processing (deseasonalizing, standardizing, etc) is performed out of sample, though in practice this can be challenging due to lack of independent data samples.

4. Data, workflows and analyses should be transparent and easily reproducible across different big-volume datasets. This can technically be achieved by linking open-source softwares to big climate data platforms, then studies should provide access to the source code, the actual AI model (via appropriate repository, e.g. Github) and exact data used, including pre-processing and post-processing (on publicly accessible data platforms, e.g. Climate Data Store).

5. Studies should use standardized benchmark datasets and multiple skill-metrics. The use of single and/or uncritical skill metrics (e.g. correlation or area under the ROC curve) can easily lead to inflated skill estimates.

## 5. FUTURE PERSPECTIVES

Artificial intelligence has proven powerful to target potential windows of predictability and eventually improve forecast skill at subseasonal to decadal time scale. In the last five years, the presence of machine/deep learning algorithms has exponentially grown in studies targeting the prediction of extreme events weeks and, to a lesser extent, seasons ahead. Evolution and progress in this topic has been extraordinarily quick, and we expect even faster-growing development.

For predictions on inter-annual to decadal time scales, the observational record provides relatively few independent samples, which impedes robust training and therefore the proliferation of studies on extremes at such a time-scale. To be applicable for predictions of the real-world climate, this requires that sufficient useful information can be learnt from the numerical model simulations (see section 4.1). Climate prediction applications of such transfer learning implementations include IOD seasonal predictions (Ling et al. 2022), Medium-Range Weather Forecasts (Rasp et al. 2021) or reconstruction of climate observations (Kadow et al. 2020). The aspect of increasing the training sample size makes this a promising approach also for climate predictions, in particular on longer (inter-annual and beyond) time scales.

Recent work has developed methods to constrain or sub-select simulations from large climate model ensembles in order to improve the skill of these decadal and even multi-decadal climate projections by aligning the phases of internal variability modes with the observed climate (Mahmood et al. 2022; De Luca et al., 2023). These constraints involve a large number of choices to be made, which imply sensitivities of the results to specific prediction targets both in space and time. We suggest that ML/DL can also be useful to further optimise these methods, e.g. by learning the most effective constraining criteria or identifying optimal analogues to select those simulations providing the highest skill at a specific region and point in time.

The main problems related to training linger on the representativeness and comprehensiveness of data, therefore if the ML/DL model can be generalized, and on the boundary between being physical-knowledge-driven and data-driven (Balaji, 2021). Dispensing with the underlying structure of equations seems impossible, unless one wants to face a number of issues. As the learning is only as good as the training data, one may find that the resulting ANN violates some basic physics (such as conservation laws), or does not generalize well. In hydrological modeling, these issues have been addressed by employing an end-to-end hybrid modeling approach based on a ML/DL algorithm constrained by energy (Zhao et al., 2019) or water (Kraft et al., 2022) conservation. Beucler et al. (2021) conducted a successful experiment where they emulated convective processes using NN while enforcing conservation laws. So-called physics-constrained machine learning thrived at extracting the information in observations, while maintaining model interpretability and physical consistency. Ideally, one could go further and venture into learning the fundamental physics. There have been attempts to learn the underlying equations for well-known systems (Brunton et al., 2016) and efforts to learn the structure of parameterizations from data, with the advantage of intrinsic interpretability (Zanna and Bolton, 2020). In this context, 'learning the physics' means writing down closed-form equations for the effect of unresolved physics on resolved-scale tendencies, using the relations learnt by the data. Resolving the turbulent vertical mixing in the atmospheric boundary layer, for example, may be key to the full understanding of the atmosphere-land process able to modulate heat waves at the subseasonal and seasonal scales.

As pointed out in the sidebar, fully exploiting the potential of those techniques requires addressing issues related to trust in the AI models. Introducing open-source benchmark datasets can enhance the community's confidence in the use of AI, by providing a framework to compare different models on common grounds (O et al., 2020; Mamalakis et al. 2022b; see also section 4.5). Yet, similar examples for climate prediction of extreme events are not available, and their production is strongly encouraged. In particular, introducing a platform to ensure the reproducibility of experiment according to the FAIR (Findable, Accessible, Interoperable, Reusable; Wilkinson et al., 2016) approach towards ML applications in weather and climate appears undelayable. The Canonical Workflow Framework for Research (CWFR) has been recently proposed to ensure the FAIRness and reproducibility of these practices (Mozaffari et al., 2022), targeting not only data, but also algorithms, tools and workflows that lead to data.

Most AI-based climate prediction models developed to date provide deterministic predictions. However, as discussed in Section 4, providing predictions in a probabilistic framework can be beneficial for robust estimation of uncertainties and skill improvement. Compared with dynamical models, AI-based models provide a larger spectrum of approaches to predict probabilities, but well-calibrated estimates of uncertainties should be ensured to guarantee consistency between predicted distributions and observed frequencies. Some approaches are similar to the ones used for numerical models, such as introducing perturbations in the initial conditions or creating model ensembles. Calibration can be better achieved when methods are directly trained to output distributions, and when probabilistic loss is used. Methods to do this range from distributional regression (networks) (e.g. Schulz and Lerch, 2022, Hu et al. 2023), to (implicit) quantile networks (e.g. Bremnes 2020, Dabney et al. 2018), to histogram-estimation networks (e.g. Scheuerer 2020) or Generalised Linear Models (GLMs) and Generalised Additive Models (GAMs) (e.g. Tuel & Martius 2022).

Additionally, numerous unexplored probabilistic ML methods exist that might suit the prediction of climate extremes. Many model families, like Bayesian NNs (Polson & Sokolov, 2017) or non-parametric models like Gaussian Processes (Rasmussen & Williams, 2005), have yet to receive much attention in existing research. Likewise, the category of generative models, including Variational Auto Encoders (VAEs; Kingma & Welling, 2019), Generative Adversarial Networks (GANs; Goodfellow, 2016),

Normalizing Flows models (Papamakarios et al., 2019), or Diffusion Models (Yang et al., 2022), has also yet to be explored in the context of climate predictions. Furthermore, conformal prediction (Vovk et al., 2005) is another interesting probabilistic approach that focuses on distribution-free uncertainty quantification and reliable probabilities, which are essential to decision-making applications of climate predictions.

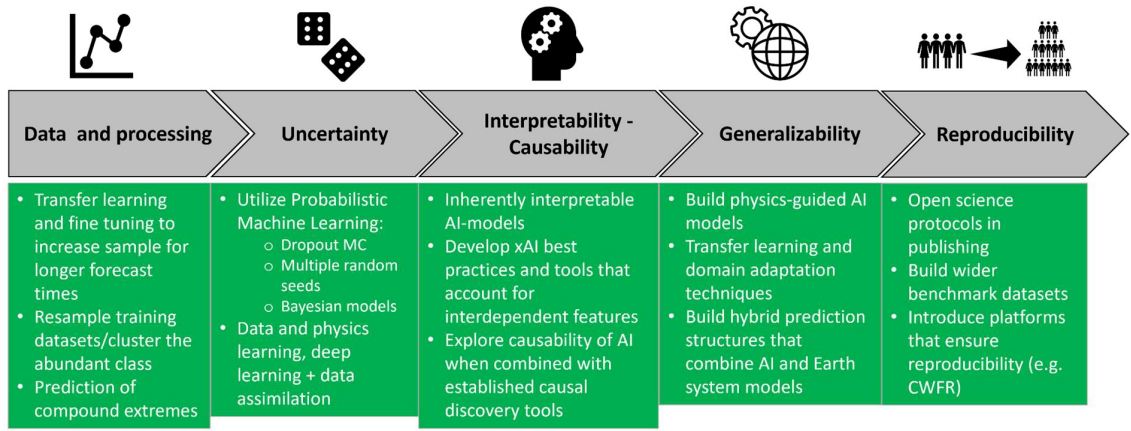

Figure 4. Perspectives and opportunities in AI-based prediction of extremes.

While AI research has been used to examine individual extreme events (see section 2), work on predictions of compound extremes is still in its infancy (Zhang et al., 2021). To the knowledge of the authors, there are hardly any published studies exploiting ML/DL for predicting compound hydrometeorological extremes, at least at the S2D time-scale.

Examples are flooding provoked by co-occurrence of high sea level and precipitation, able to cause substantial runoff in coastal areas (Bevacqua et al., 2019; Wahl et al., 2015); compound hot-dry events usually linked to persistent anticyclonic weather systems (Bevacqua et al., 2022; De Luca & Donat, 2023, Yin et al., 2023); heavy precipitation-high wind speed events that can occur during cyclonic weather (De Luca et al., 2020; Martius et al., 2016; Zscheischler et al., 2021).

At the time of writing this article, a few studies have investigated the use of AI for compound extreme events prediction, but hardly at the S2D scale. Park & Lee (2020) assessed coastal flooding as the compounding effect of high tides and heavy rainfall in South Korea, and developed a future (2030-2080) compound risk map using ML algorithms such as k-nearest neighbour, RF and support vector machine, to compensate for the shortfalls of each one individually. Sampurno et al. (2022) used an hydrodynamic model trained with a similar set of ML models, to perform compound flooding predictions over the Kapuas River delta (Indonesia) at the weather time-scale. Kondylatos et al. (2022) used DL algorithms (LSTM and ConvLSTM) to predict wildfires in the Eastern Mediterranean a few days ahead, and their XAI algorithm allowed the models to disentangle between dryness-driven and wind-driven fires. In addition, AI-based techniques already help enable the quantification of relationships between the extremes of two variables (Zhang et al., 2021). Bayesian network (Sanuy et al., 2020) and ANN (Huang et al., 2021; Feng et al., 2021) have been used to understand compound extremes. In addition, complex networks are capable of driving the casual relationship between two or more variables (Sun et al., 2018).

In conclusion, the scientific knowledge for AI-based skillful predictions of compound extremes is already available. Since co-occurrence and interaction of climate extremes often generates a compound effect resulting in more severe socio-economic impacts than the extremes taken singularly (Zscheischler et al., 2018), implementing this knowledge at the S2D time-scale may provide stakeholders with useful tools for planning climate adaptation strategies.

As the community continues to refine AI technologies and leverage their potential, we stand to gain invaluable insights that can empower us to prepare, mitigate, and adapt to climate extremes with greater precision and foresight. Ultimately, the integration of AI into climate prediction of extremes holds immense potential for building more resilient and sustainable societies in the face of an increasingly variable and changing climate.

**Funding Information**

SM has received funding from the European Union's Horizon 2020 research and innovation program under the Marie Skłodowska-Curie grant agreement No. 101033654 (ARTIST).

The research of LC was supported by the European Union's Horizon 2020 Marie Sklodowska-Curie grant agreement No. 101065985 (CYCLOPS).

PDL has received funding from the European Union's Horizon Europe research and innovation program under the Marie Skłodowska-Curie grant agreement No. 101059659

LP and MD have received funding from the European Space Agency's project AI4Drought under the AI4SCIENCE call contract No. 4000137110/22/I-EF.

**Acknowledgments**

The authors thank Antonia Frangeskou for her technical support in drawing figure 1.